\documentclass[twocolumn]{revtex4-2}

\usepackage[utf8]{inputenc}
\usepackage{graphicx}
\usepackage{xcolor}
\usepackage{soul}

\usepackage[mathlines]{lineno}
\usepackage[normalem]{ulem}
 \usepackage{amssymb,amsthm,amsmath}

\begin{document}


\title{Random search with resetting in heterogeneous environments}

\author{Luiz Menon Jr.$^{1}$,  and Celia Anteneodo$^{1,2}$}
\address{$^{1}$ Department of Physics, PUC-Rio, Rua Marqu\^es de S\~ao Vicente 225, 22451-900, Rio de Janeiro, RJ, Brazil}
\address{$^{2}$ Institute of Science and Technology for Complex Systems, INCT-SC, Brazil}

\begin{abstract}

We investigate random searches under stochastic position resetting at rate $r$, in a bounded 1D environment with space-dependent diffusivity $D(x)$. 
For arbitrary shapes of $D(x)$ and prescriptions of the associated multiplicative stochastic process, 
we obtain analytical expressions for the average time $T$ for reaching the target (mean first-passage time), given the initial and reset positions, in good agreement with  stochastic simulations.  
For arbitrary $D(x)$, we obtain an exact closed-form expression for $T$, within Stratonovich scenario, 
while for other prescriptions, like  It\^o and  anti-It\^o, we derive asymptotic approximations for small and large rates $r$. Exact results are also obtained for particular forms of $D(x)$, such as the linear one, with arbitrary prescriptions,  allowing to outline and discuss the main effects introduced by diffusive heterogeneity on a random search with resetting.  
We explore how the effectiveness of resetting varies with different types of heterogeneity.

\end{abstract}

 \maketitle

\section{Introduction}
\label{sec:introduction}

The topic of random searches is relevant across a range of scales, from molecular interactions like protein-DNA binding~\cite{mirny2009protein, chen2019target, bhattacherjee2014search}, to ecological contexts where finding food and viable habitats is crucial for survival~\cite{LuzNature1999, o1990search, bartumeus2005animal, viswanathan2011physics, MartinezGarcia2013, Redner2022}. Studies in this field can provide effective strategies for exploring complex spaces and identifying targets~\cite{chupeau2015}, with potential technological applications, as in robotic design~\cite{castello2016} or in the optimization of  neural networks~\cite{bergstra2012}.
 
The introduction of stochastic resetting in random search problems has gained significant attention over the last decade \cite{evans2011diffusion}. 
This strategy, involving the resetting of the state or position of a stochastic dynamics can  significantly decrease the average time required to reach a target or boundary. It has been shown to accelerate molecular dynamics calculations~\cite{blumer2024combining} and be used to optimize complex algorithms~\cite{montanari2002optimizing}.
This is because, by resetting, 
a stochastic trajectory has a chance to escape local traps or unfavorable regions, thus improving the efficiency in completing a search task, for a suitable restart rate~\cite{reuveni16}. 
There are regions in parameter space where resetting can increase  the  mean time to reach the target, or more generally the mean first-passage time (MFPT), indicating that  resetting can hinder rather than facilitate the search process.
But, in other cases, resetting can be favorable, 
existing an optimal resetting rate that minimizes the average time to reach the target in a bounded domain~\cite{christou2015diffusion}. 
Recent advances highlighted the benefits of resetting strategies in bounded domains~\cite{durang2019first}. Even when a random walker position is reset  to a place farther from the target than its original location, resetting can still improve the search by decreasing the MFPT.

In these studies, the environment typically has uniform diffusivity, while in many real cases the diffusion coefficient may be space-dependent. 
This type of heterogeneous media appears in  the description of diverse physical complex systems presenting anomalous diffusion, e.g., subdiffusion~\cite{zodage2023sluggish}
or turbulent diffusion~\cite{stella2022}, as well as   in the modeling of 
 biological~\cite{english2011single,neuralPRX}, 
 socioeconomic~\cite{vieira2018threshold}, 
and ecological \cite{dos2020critical} systems, among other examples. 
In the context of random searches, 
  even in the absence of resetting, 
studies about search processes in heterogeneous environments, involving position-dependent diffusion coefficient, are relatively few and recent~\cite{vaccario2015first,godec2015optimization,mutothya2021first,li2020particle,Santos2022}. 
There are previous works which deal with   heterogeneous diffusion under resetting in semi-infinite environments and for particular forms of the diffusitivity~\cite{lenzi2022transient,Ray2020}, but as far as we know, results  in bounded domains, for more general forms of heterogeneity, are still lacking.

Then, we address the problem of search with resetting in a confined (one-dimensional) environment with a space-dependent diffusion coefficient $D(x)$. 
The dynamics of the searcher     
follows a Langevin dynamics and 
with rate $r$ is reset to a position $x_r$, namely
\begin{eqnarray}
x(t+\delta t) =\left\{ \begin{array}{cl} 
    \stackrel{ 1- r \delta t}{=}  &  x(t) + \sqrt{2D(x^*)}\eta(t) \delta t,   \\
    \stackrel{ r \delta t}{=} & x_r  ,
\end{array}   
\right.
\label{eq:process1}
\end{eqnarray}
where $\eta(x)$ represents a delta-correlated Gaussian noise and   $D(x)>0$. 
Importantly, since the white noise is multiplicative, it is necessary to choose the specific instant at which $x^*$ is calculated.  
We consider $x^*=  [(2-A)\,x(t+dt)+A\,x(t)]/2$~\cite{vaccario2015first}, where $A \in[0,2]$, focusing on three special interpretations:  Itô ($A=2$)~\cite{ito1944109},  
Stratonovich ($A=1$)~\cite{stratonovich1966new}, anti-Itô  ($A=0$)~\cite{hanggi1982nonlinear,Klimontovich1990}. 
In the It\^o form, Eq.~(\ref{eq:process1}) can be cast as  
\begin{align}
x(t+\delta t) =\left\{ \begin{array}{cl} 
    \stackrel{ 1- r \delta t}{=}  &  x + \frac{2-A}{2} D'(x) + \sqrt{2D(x)\delta t }\,\eta(t)   \\
    \stackrel{ r \delta t}{=} & x_r \,,
\end{array}   
\right.
\label{eq:process1b}
\end{align}
where $x=x(t)$, $D'=dD/dx$   rules  the  noise-induced drift~\cite{volpe2016effective}. 
We consider that the target is located at $x_t=0$ (represented by an absorbing boundary) while we set a reflecting boundary at $x=L$. 

The Fokker-Planck equation associated to Eq.~(\ref{eq:process1b}), joining previous results for heterogeneous diffusion~\cite{,PhysRevE.99.042138} and stochastic resetting~\cite{evans2011diffusion}, is given by %
\begin{align}
\frac{\partial \ }{\partial t} p(x,t|x_0)&= \frac{\partial \ }{\partial x}\left\{ D(x)^{1-\frac{A}{2}} \frac{\partial}{\partial x}  [D(x)^{\frac{A}{2}}p(x,t|x_0) ]      \right\}\nonumber \\ &-  r p(x,t |x_0) + r \delta(x-x_r) ,
\label{eq:FPEall}
\end{align}
where  $x_0,\; x_r\in (0,L]$, plus the given boundary conditions at the borders of the interval.

To study first-passage statistics, we follow the usual approach, which is to consider the backward Fokker-Planck equation, where $x_0$ is a variable, namely
\begin{align}
\frac{\partial}{\partial t} Q(x_0,t) &= D(x_0)^{\frac{A}{2}} \frac{\partial}{\partial x_0} \left\{ D(x_0)^{1-\frac{A}{2}} \frac{\partial}{\partial x_0} Q(x_0,t) \right\} \nonumber \\&+ rQ(x_r,t) - rQ(x_0,t), \label{eq:backwardA}
\end{align}
where $Q(x_0,t ) = \int_0^L p(x,t|x_0)dx$  is the survival probability. 
Applying the Laplace transform (denoted by a tilde) to Eq. (\ref{eq:backwardA}) and considering that 
$T(x_0)=\tilde{Q}(x_0, s \to 0)$~\cite{redner2001guide},   we obtain
\begin{equation}
     D(x_0)^{\frac{A}{2}} \left[ D(x_0)^{1-\frac{A}{2}} T'(x_0) \right]' + rT(x_r) - rT(x_0)=-1.
    \label{eq:mfpt-general}
\end{equation}
This ordinary differential equation, plus boundary conditions, will be the starting point for our analytical derivations of the MFPT, which is a measure of the efficiency of the search.

The paper is organized as follows. 
In Section \ref{sec:linear}, we present both analytical and simulation results for a linear space-dependent diffusion coefficient, and arbitrary $A$, which gives insights on the effect of heterogeneity, given by $D(x)$ and $A$. 
In section \ref{sec:arbitrary Dx}, we deal with arbitrary profiles, for which we obtain closed expressions of the MFPT within  Stratonovich interpretation and asymptotic approximations  for arbitrary $A$, allowing to discuss the impact of non-monotonic profiles on the efficiency of random search under resetting. 
Although derivations are performed for arbitrary values of the parameters, we will focus on the cases where the reset position coincides with the initial one, i.e., we will set $x_r=x_0$.
Other situations, such as,  
environments with two absorbing boundaries, 
the limit of semi-infinite domains, 
and $x_r\neq x_0$ are discussed in section \ref{sec:other}. 
Final remarks are presented in section.~\ref{sec:final}.

\section{Linear diffusivity profiles with arbitrary $A$}
\label{sec:linear}

Let us first consider a simple and manageable class of heterogeneous diffusivity profiles, with the linear form
\begin{equation} \label{eq:linearD}
D(x) =D_0+ \alpha(x -L/2),
\end{equation}
where $\alpha$ is a real parameter subject to the condition $|\alpha|<2D_0/L$ to warrant positivity of $D(x)$, and  $D_0$ is the average diffusivity,  the same for all $\alpha$, to make  comparisons fair.

We start by substituting Eq.~(\ref{eq:linearD}) into  Eq.~(\ref{eq:mfpt-general}), yielding
\begin{equation}
    T''(y_0) + \frac{1-A}{y_0+\lambda}T'(y_0) + r T(y_r) - rT(y_0)=-1,
    \label{eq:edofpt}
\end{equation}
where $\lambda = 2 \sqrt{D_0 - \alpha L/2 }/\alpha$, and we used the change of variables 
\begin{equation} \label{eq:y}
y(x)=\int_0^x D(x')^{-1/2}dx',     
\end{equation}  
which in the present case explicitly becomes
\begin{equation}
y(x) = \frac{2}{\alpha} \left(\sqrt{D_0 + \alpha[x-L/2]}- \sqrt{D_0-\alpha L/2} \right).
\end{equation}
The details for solving Eq.~(\ref{eq:edofpt}), 
with a reflecting boundary at $x=L$, 
are presented in Appendix~\ref{app:mfpt-linear}. 
The particular solution  reads
\begin{eqnarray}
    T(y_0)&=&  \frac{
 \left(\frac{z_t}{z_r}\right)^{\frac{A}{2}} U_A(z_t,z_L)
-\left(\frac{z_0}{z_r}\right)^{\frac{A}{2}} U_A(z_0,z_L) 
}{r U_A(z_r,z_L)}, \hspace*{5mm} \label{eq:mfpt-r-L}
\end{eqnarray}
where we defined
$x_L\equiv L$, $z_* \equiv \sqrt{r}(y(x_*)+\lambda)$.    
and
\begin{eqnarray}
\label{eq:U_def}
    U_A( v,w) \equiv  K_{\frac{A}{2}}\left( v \right)I_{\frac{A-2}{2}}(w) + I_{\frac{A}{2}}(v)K_{\frac{A-2}{2}}(w),
\end{eqnarray}
being $I$ and $K$   the modified Bessel functions.

Note that, in the limit  $\alpha\to 0$, $y(x) = x/\sqrt{D_0} $, and $\lambda$ goes to infinity, recovering the known result for the MFPT of the homogeneous case~\cite{durang2019first}, which is 
\begin{equation}
    \label{eq:homo_reflect}
    T(x_0) = \frac{\cosh\left[\sqrt{r/D_0}\,L \right] - \cosh \left[\sqrt{r/D_0}\,(x_0 -L) \right] }{r\cosh \left[\sqrt{r/D_0} (L-x_r)\right]}.
\end{equation}

\begin{figure}[hb!]
    \centering
\includegraphics[width=0.45\textwidth]{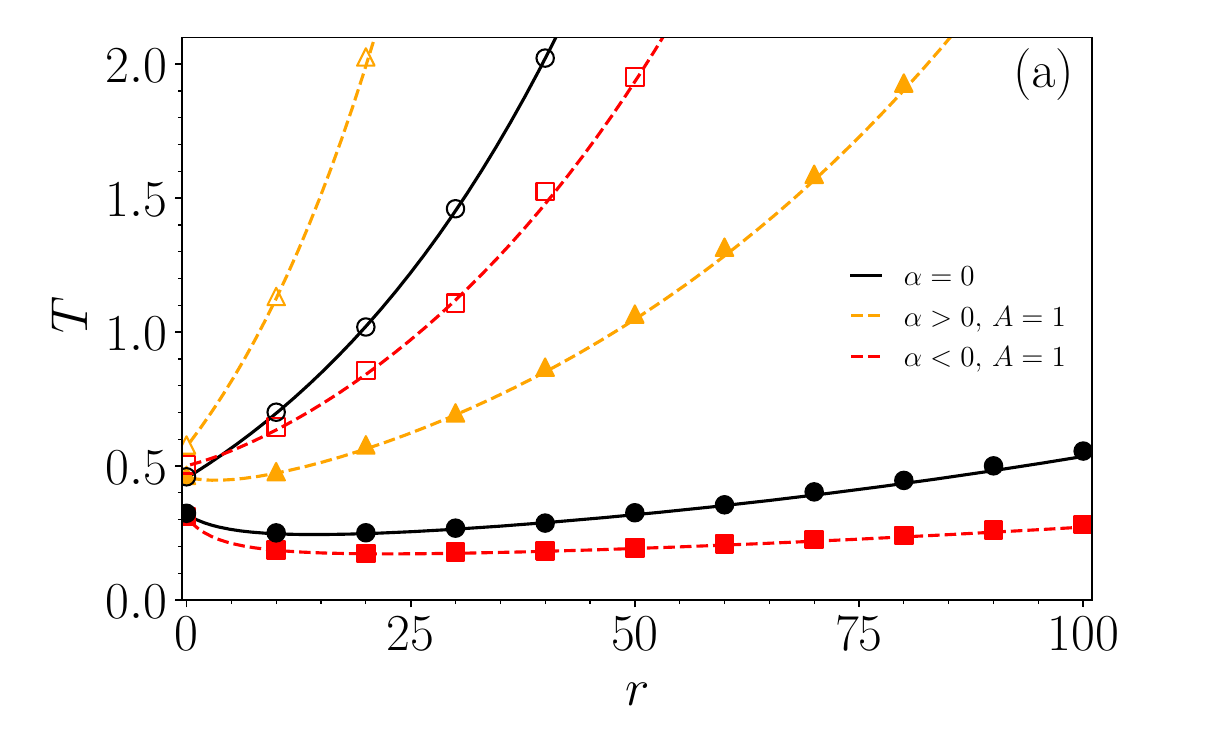}
\includegraphics[width=0.45\textwidth]{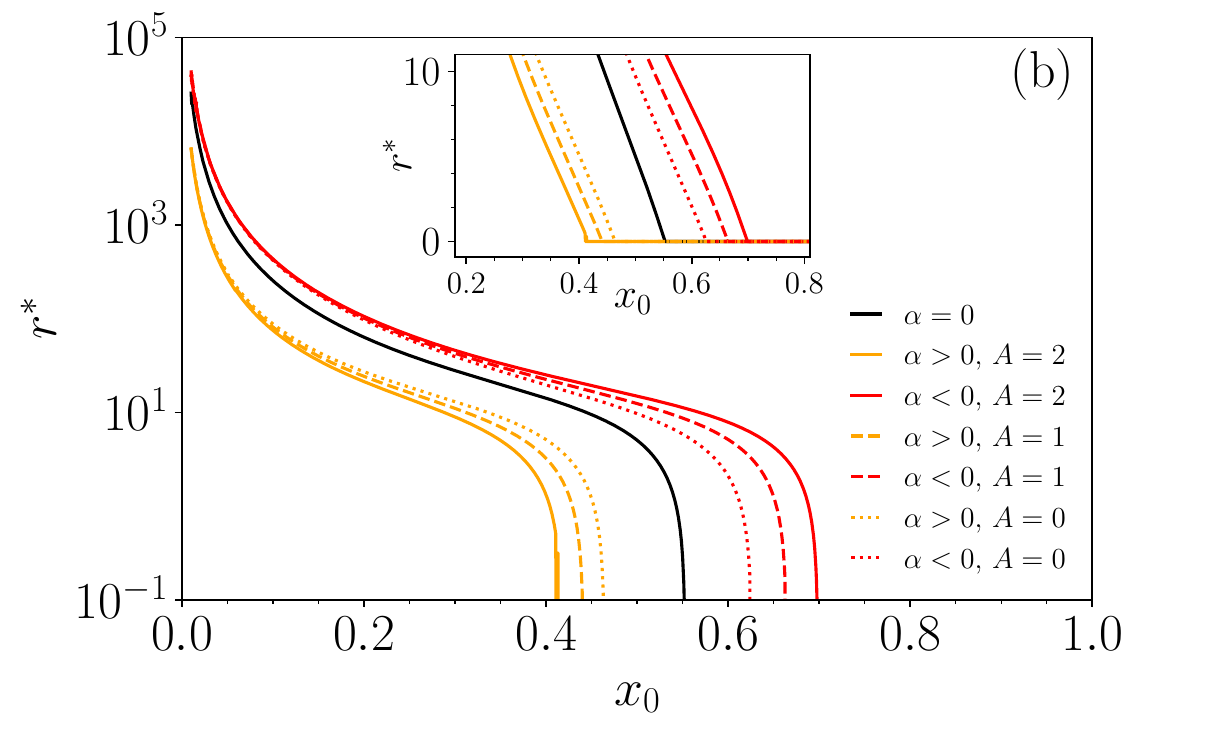}
\includegraphics[width=0.45\textwidth]{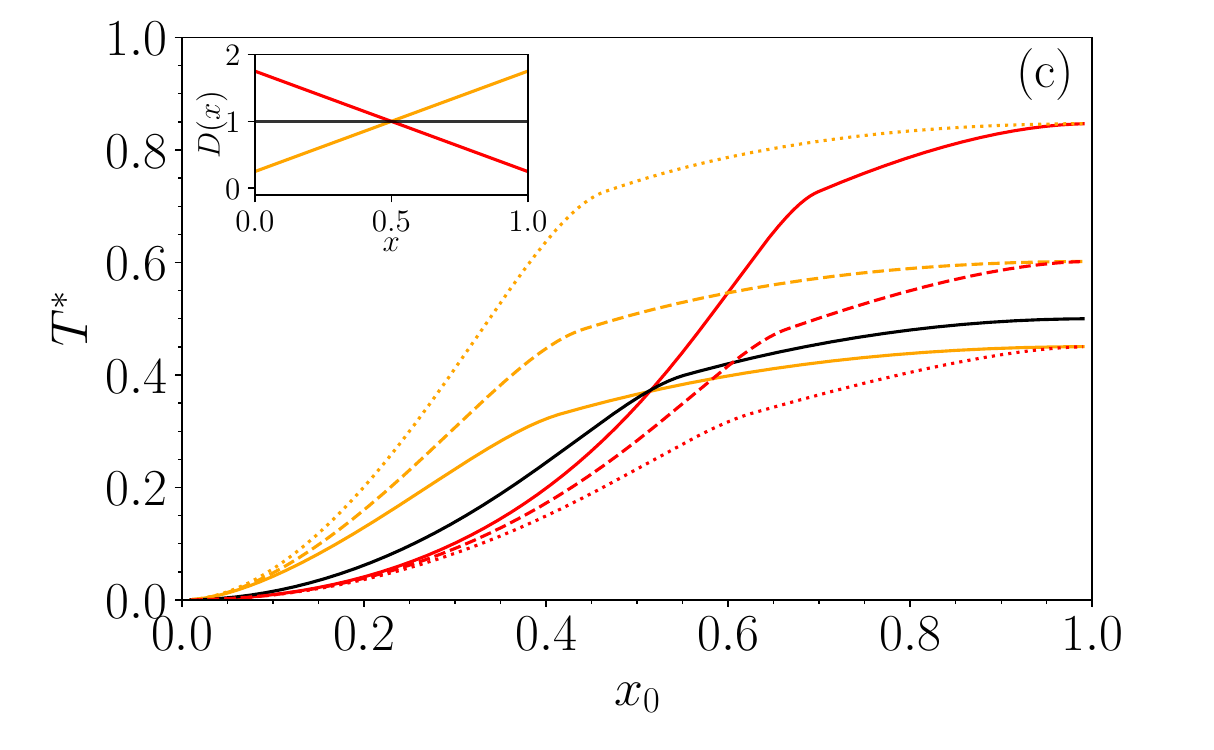}
    \caption{
    {\bf Linear diffusivity $D(x)=1+\alpha(x-1/2)$,} 
       with $\alpha = + 1.5$ (light orange), $\alpha =  -1.5$ (red), and $\alpha=0$ (black), depicted in the inset of the lower panel. 
    (a)  $T$ vs. $r$, provided by Eq.~(\ref{eq:mfpt-r-L}), for $x_0 = 0.4$ (filled symbols),  and $x_0 = 0.7$ (hollow), considering the Stratonovich interpretation ($A=1$). 
    (b) Optimal resetting rate $r^*$ (that minimizes the MFPT) vs. $x_0$, for different values of $A$ indicated in the legend. 
    The inset shows a magnification of the critical region in linear scale. 
    (c) Corresponding optimal MFPT $T^*$ vs. $x_0$.
    Symbols correspond to numerical simulations (average over $10^4$ trajectories) of Eq.~(\ref{eq:process1b}), and lines to theoretical results.
    }
    \label{fig:linearD}
\end{figure}

The theoretical prediction of $T$ as a function of the resetting rate $r$, according to  Eq.~(\ref{eq:mfpt-r-L}) together with Eq.~(\ref{eq:y}), 
is depicted in Fig.~\ref{fig:linearD}(a), 
for two reset positions $x_0$ (recalling  $x_r=x_0$), and for slopes $\alpha = -1.5$ (decreasing), 0 (constant) and 1.5 (increasing), within Stratonovich interpretation ($A=1$). 
Moreover, we set  $L=1$ and $D_0=1$ in this and other examples. 
Two different regimes emerge. 
For large enough reset position $x_0$ (e.g., $x_0=0.7$ in the figure), the MFPT increases monotonically with $r$ for any $\alpha$. Therefore,  in this case stochastic resetting does not favor the search. 
Otherwise (as illustrated by $x_0=0.4$), 
the MFPT has a minimum, $T^*$, indicating the existence of an optimal non-null value of $r$, $r^*$, that minimizes the MFPT. 
These two regimes occur  also for $A \neq 1$, 
as well as in the homogeneous case $\alpha=0$ (black line), as previously reported~\cite{christou2015diffusion, durang2019first}, 
but, as we will see, the type of heterogeneity (given by $D(x)$ and $A$) influences the critical $x_0$, as well as the optimal point $(r^*,T^*)$, modifying the efficiency of the search.

In Fig \ref{fig:linearD}($b$), we show $r^*$ vs. $x_0$, for the same profiles,  
with  $A=0,1,2$. 
The interval of $x_0$ where $ r^* \neq 0 $ indicates the regime where the rate $r^* $ minimizes the MFPT, while  $r^* = 0$  corresponds to the regime where resetting  hinders the search.
Note that for $\alpha<0$ (red lines),  
  the critical value $x_{c}$ between both regimes is larger than for the homogeneous case with the average diffusivity $D_0$ for any $A$, more pronouncedly the larger $A$ (the less anticipating the random walk). 
 This means that the range of reset positions $x_0$ for which  resetting mechanism  optimizes  the search is wider,  compared to the homogeneous case.  
 Therefore, resetting is effective for positions further away from the target than in the homogeneous case.
Opposite tendencies occur for increasing profiles ($\alpha>0$, light orange lines). 

The corresponding optimal times $T^*$ vs.  $x_0$ are shown in Fig.~\ref{fig:linearD}($c$). 
As a general feature,  note that  in the 
more anticipating case ($A=0$, anti-It\^o), 
decreasing (increasing) profiles are more (less) efficient, for any $x_0$.  

For other interpretations, crossings occur for intermediate values of $x_0$.
For small $x_0$, higher efficiency is attained for $\alpha<0$ (red lines), more, the more anticipating the process.
Opposite tendencies occur for increasing profiles ($\alpha>0$, light orange lines). 

In the opposite limit of $x_0 \to L$, 
the results for $(\alpha, A)$ and $(-\alpha, 2-A)$ tend to coincide, then for ($A=1$, Stratonovich)
 increasing and decreasing profiles with same $|\alpha|$ behave equally, a symmetry property known for the case without resetting~\cite{menon2023}.  
  
Let us mention that the features remarked above  are not exclusive of the linear case but the outcomes are qualitatively similar for other monotonic forms such as $D(x)=D_0+d\cos(\pi x/L)$ or simply the two level form $D(x)=D_0\pm d$, for $x\lessgtr L/2$, (not shown).

\section{Arbitrary $D(x)$ }
\label{sec:arbitrary Dx}

 \subsection{Stratonovich scenario ($A=1)$}
\label{sec:stratonovich}

Within the Stratonovich framework, it is possible to obtain a closed expression of the MFPT for arbitrary profiles, as follows.
Setting $A=1$, Eq.~(\ref{eq:mfpt-general}) becomes
\begin{equation}
\label{eq:edo_ts}
    \sqrt{D(x_0)} \left[ \sqrt{D(x_0)} T'(x_0) \right]' + rT(x_r) - rT(x_0)=-1.
\end{equation}
Applying also in this case the  change of variables 
given by Eq.~(\ref{eq:y}), we obtain
\begin{equation}
    T''(y_0)  - r T(y_0) + rT(y_r) = -1,
\end{equation}
whose  general solution has the self-consistent form 
\begin{equation}
\label{eq:ts}
 T(y_0)= \frac{1}{r} +  T(y_r)   + \sum_{m = \pm} c_{m} \exp{\left[m\,\sqrt{r} \,y_0 \right]}, 
\end{equation}
where $c_\pm$ are determined by the boundary conditions. Then, we obtain the explicit expression  

\begin{equation} \label{eq:A=1}
    T(y_0) = \frac{
    \cosh\left[ \sqrt{r} ( y_L - y_t)   \right] - \cosh\left[ \sqrt{r} ( y_L - y_0)   \right]    }{r\, \cosh\left[ \sqrt{r} ( y_L - y_r)\right] },
\end{equation} 
which for homogeneous profiles, recovers Eq.~(\ref{eq:homo_reflect}).

\begin{figure}[ht!]
    \centering
    \includegraphics[width=0.45\textwidth]{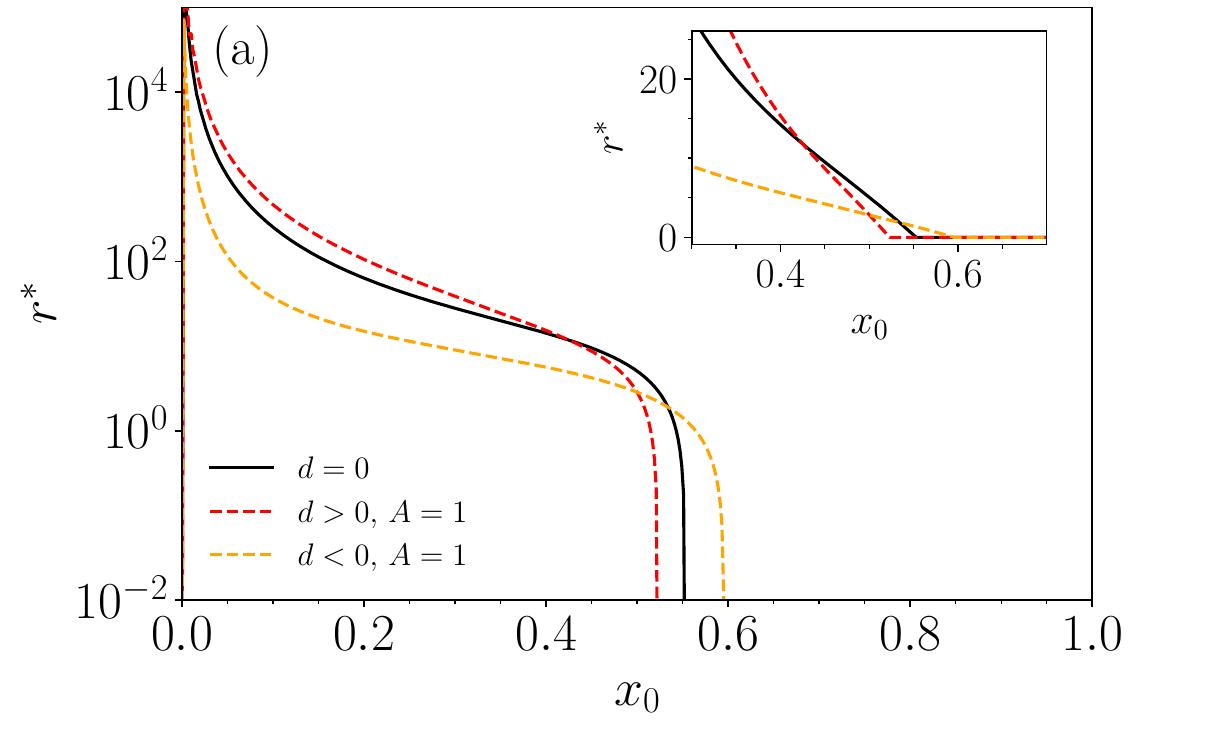}
    \includegraphics[width=0.45\textwidth]{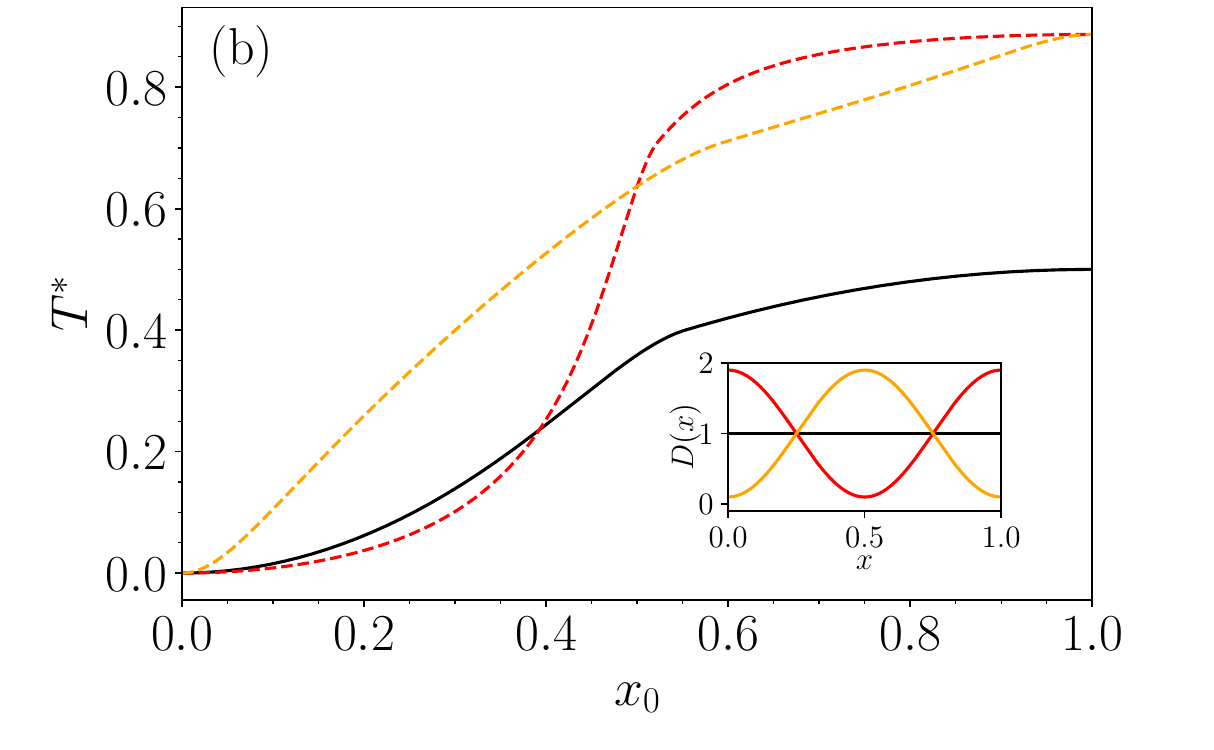}
    \caption{
      {\bf Oscillating diffusivity}  $D(x) = 1 + d\cos\left( 2\pi x \right)$, with $A=1$ , depicted in the inset of the lower panel. 
    (a) Optimal resetting rate, $r^*$,  
    and  (b) corresponding optimal MFPT, $T^*$  vs. $x_0$, for 
    $d = 0.9 $ (orange), $d =- 0.9 $ (red) and $d=0$ (black). 
    The inset in (a) is a magnification around the critical $x_0$.
    }
    \label{fig:cosk2}
\end{figure}

For  non-monotonic profiles, we consider the family
\begin{equation} \label{eq:Dcos}
D(x) = D_0 + d\cos( k\pi x /L)\,,
\end{equation}
where $|d|<D_0$ for positivity, and $k$ is integer to preserve the average $D_0$.

\begin{figure}[ht!]
    \centering
    \includegraphics[width=0.45\textwidth]{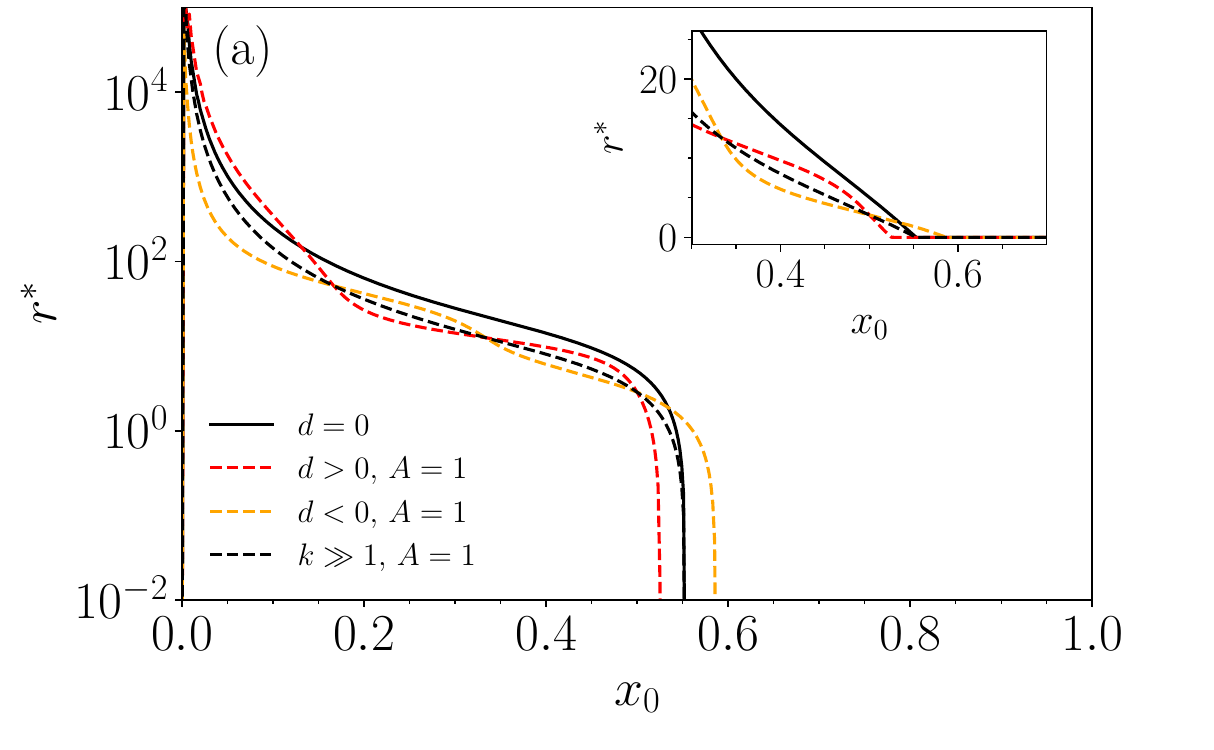}
    \includegraphics[width=0.45\textwidth]{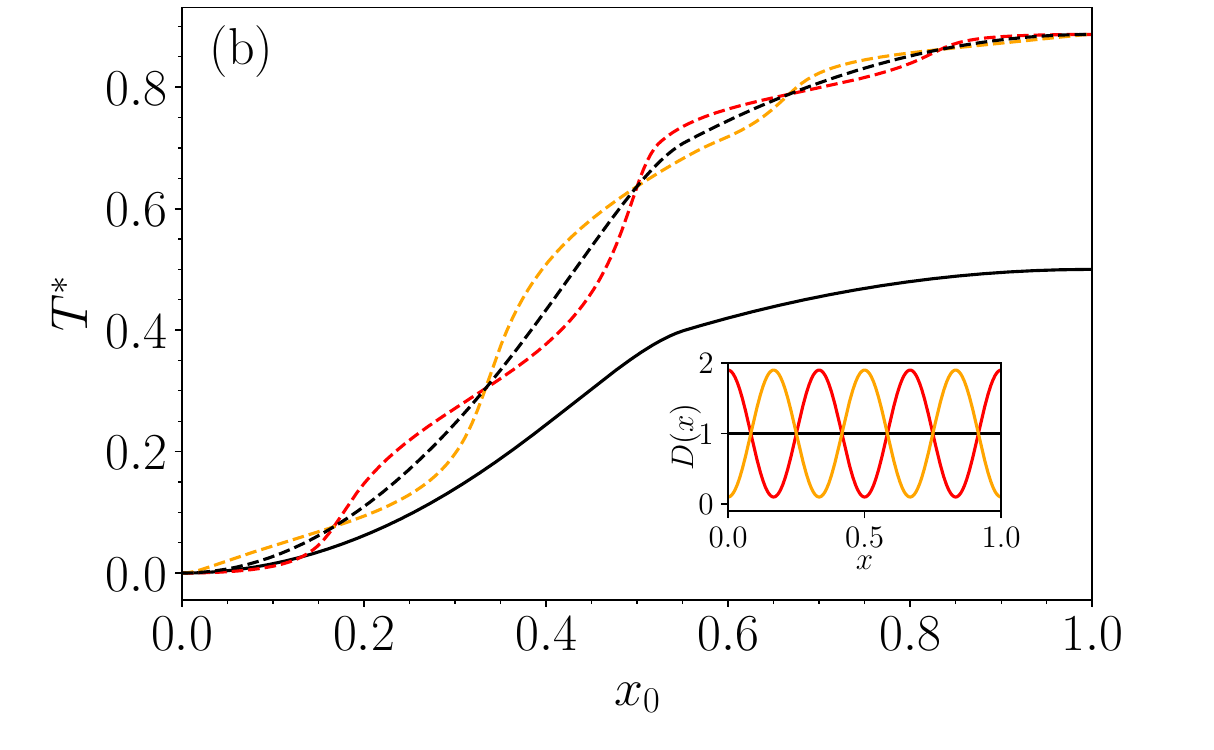}
    \caption{
    {\bf Oscillating diffusivity} $D(x) = 1 + d\cos\left( 6\pi x \right) $, with $A=1$,    depicted in the inset of the lower panel. 
    (a) Optimal resetting rate, $r^*$,  
    and  (b) corresponding optimal MFPT, $T^*$  vs. $x_0$, for 
    $d = 0.9 $ (orange), $d =- 0.9 $ (red) and $d=0$ (black).  
        The inset in (a) is a magnification around the critical $x_0$.  
        The homogeneous case is also plotted for comparison (black lines).
    }
    \label{fig:cosk6}
\end{figure}

The optimal $r^*$ and $T^*$ vs. $x_0$ 
are presented in Fig.~\ref{fig:cosk2}, for   $k=2$, $D_0=1$, $L=1$. 
In general lines, for small reset positions $x_0$, we observe similar tendencies to those in the monotonic cases, depending on  the slope of the diffusivity profile in the vicinity of the target.   
This trend still holds, but closer to the target, if we increase the number of wrinkles, as illustrated in Fig.~\ref{fig:cosk6}, for $k=6$.  
For highly oscillating profiles ($k\gg 1$), the MFPT is larger than in the homogeneous case for any reset position $x_0$, as can be seen by the dashed lines in Fig.~\ref{fig:cosk6}. Therefore, at least in the Stratonovich scenario, the search, even when resetting is beneficial, is less efficient than in the homogeneous case.

\subsection{Arbitrary $A$}
\label{sec:abritray A}

For interpretations other than Stratonovich,  we could not find a closed expression of the MFTP for arbitrary profiles. 
However, we managed to obtain approximate expressions valid in the limits of small and large resetting rate $r$, that will be useful  to determine the  type of regime (that is, if $r\neq 0$ improves the search or not) and estimate the optimal $r$. 
The  large-$r$ approximation is obtained using the WKB method ~\cite{kramers1926wellenmechanik,wentzel1926verallgemeinerung}.
The approximation valid for small $r$ is derived using standard perturbation theory \cite{butkov1973mathematical}. 
We summarize below the explicit expressions found in each case, while details of the derivations are left for Appendix \ref{app:asymptotic}. 
The good performance of these approximations is illustrated in Fig.~\ref{fig:comparison} of Appendix \ref{app:asymptotic}, for a linear profile, compared with the exact expression was obtained in Sec.~\ref{sec:linear}. 

\begin{figure*}[ht!]
\includegraphics[width=0.3\textwidth]{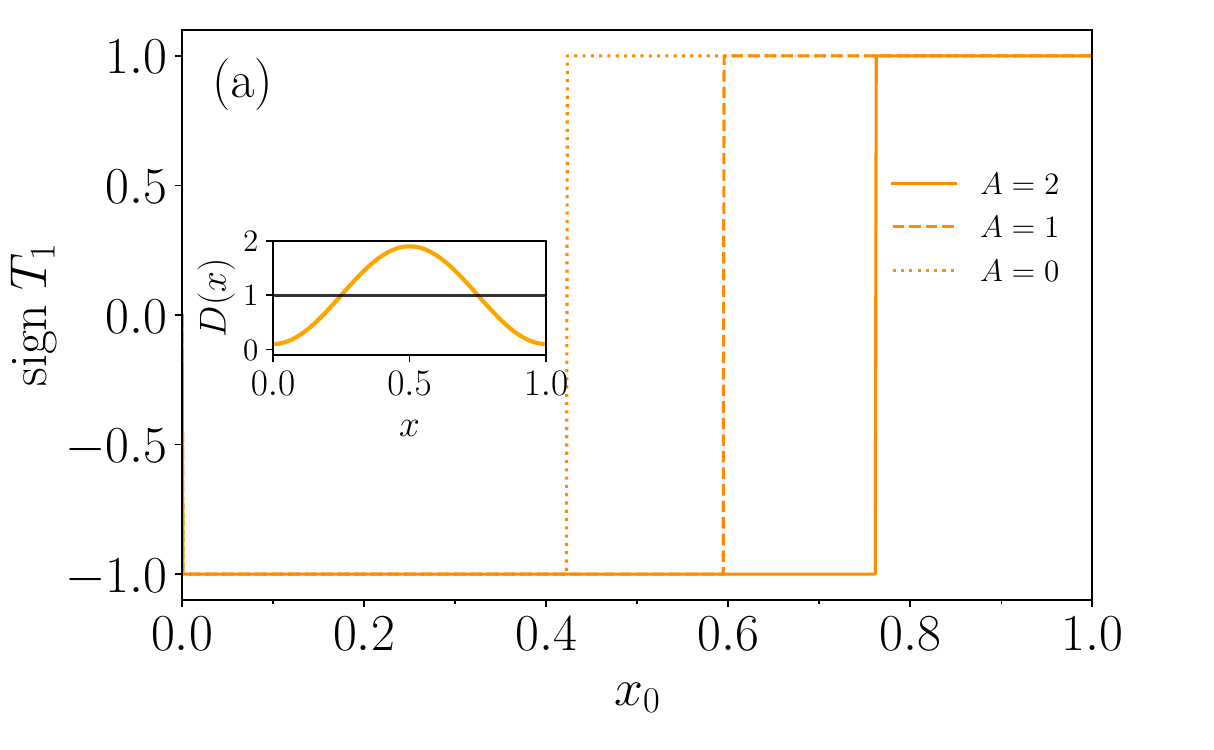}
\includegraphics[width=0.3\textwidth]{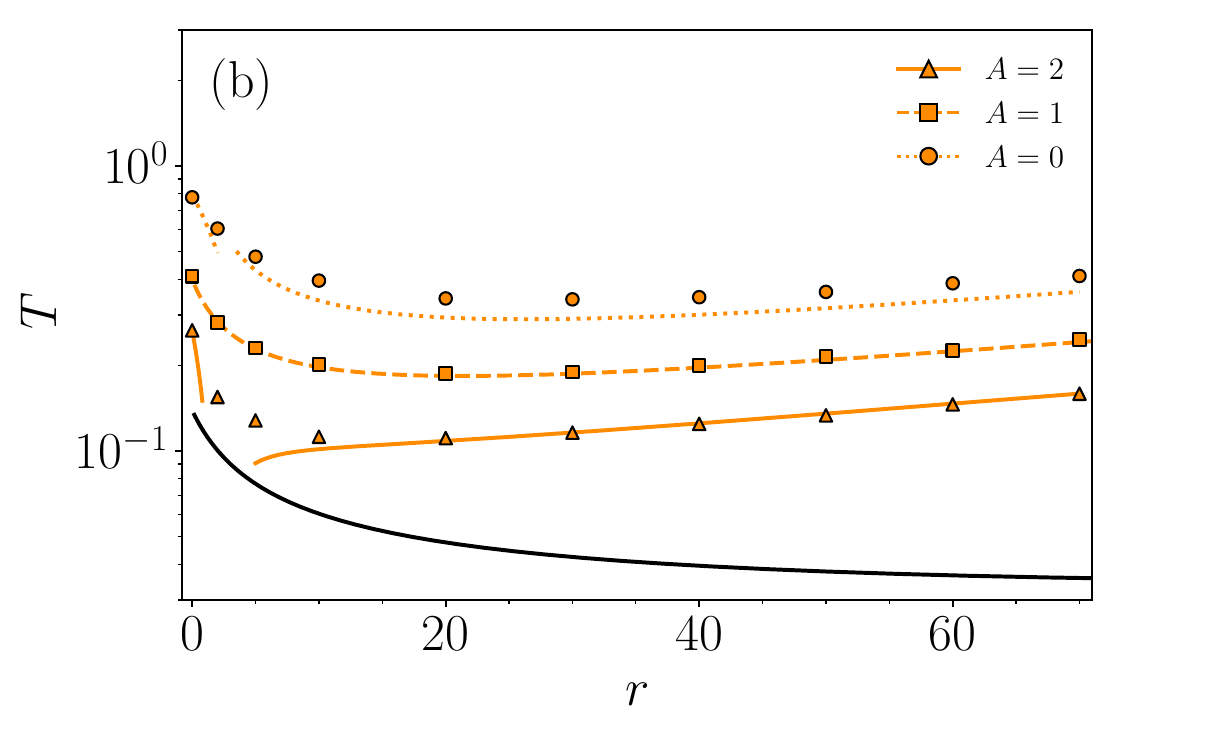}
\includegraphics[width=0.3\textwidth]{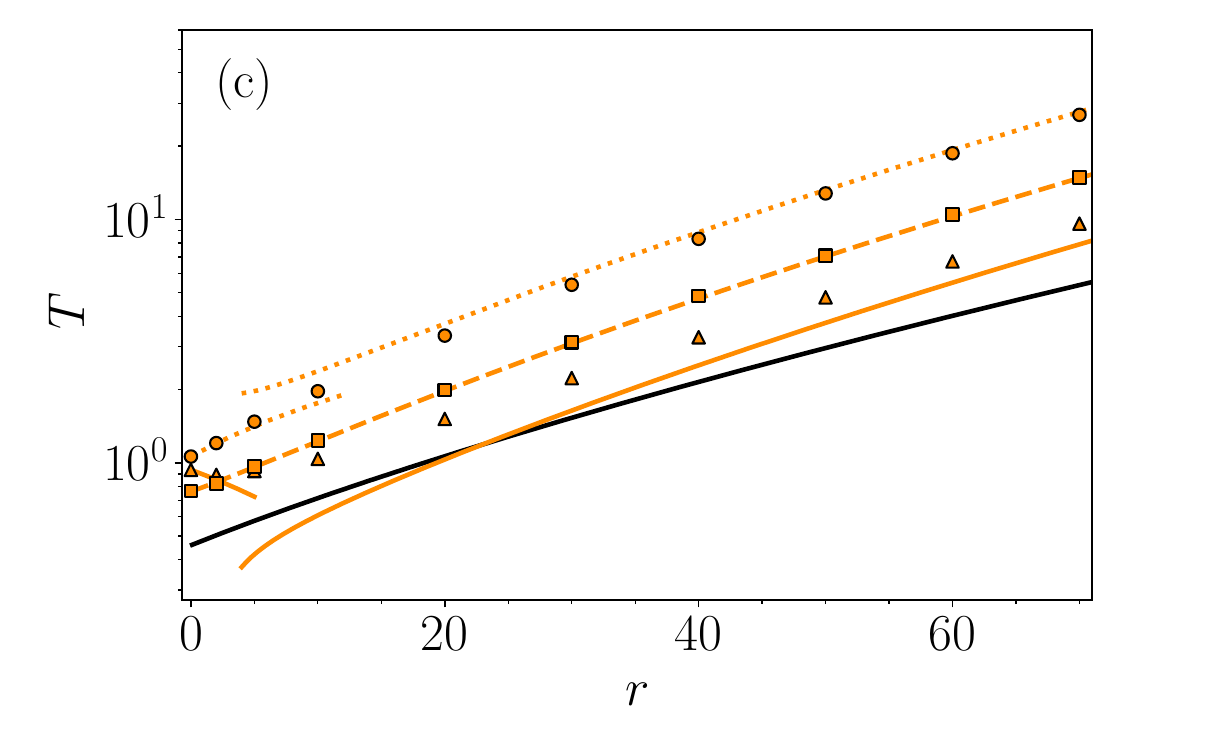}
\includegraphics[width=0.3\textwidth]{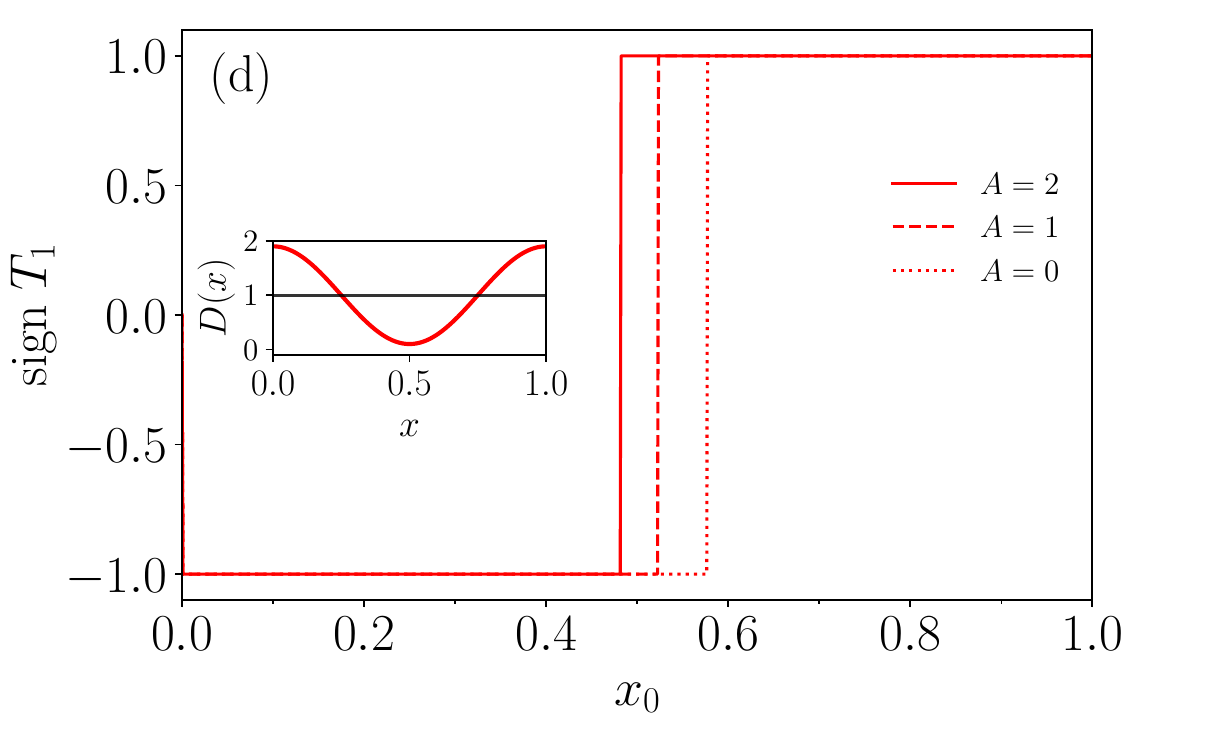}
\includegraphics[width=0.3\textwidth]{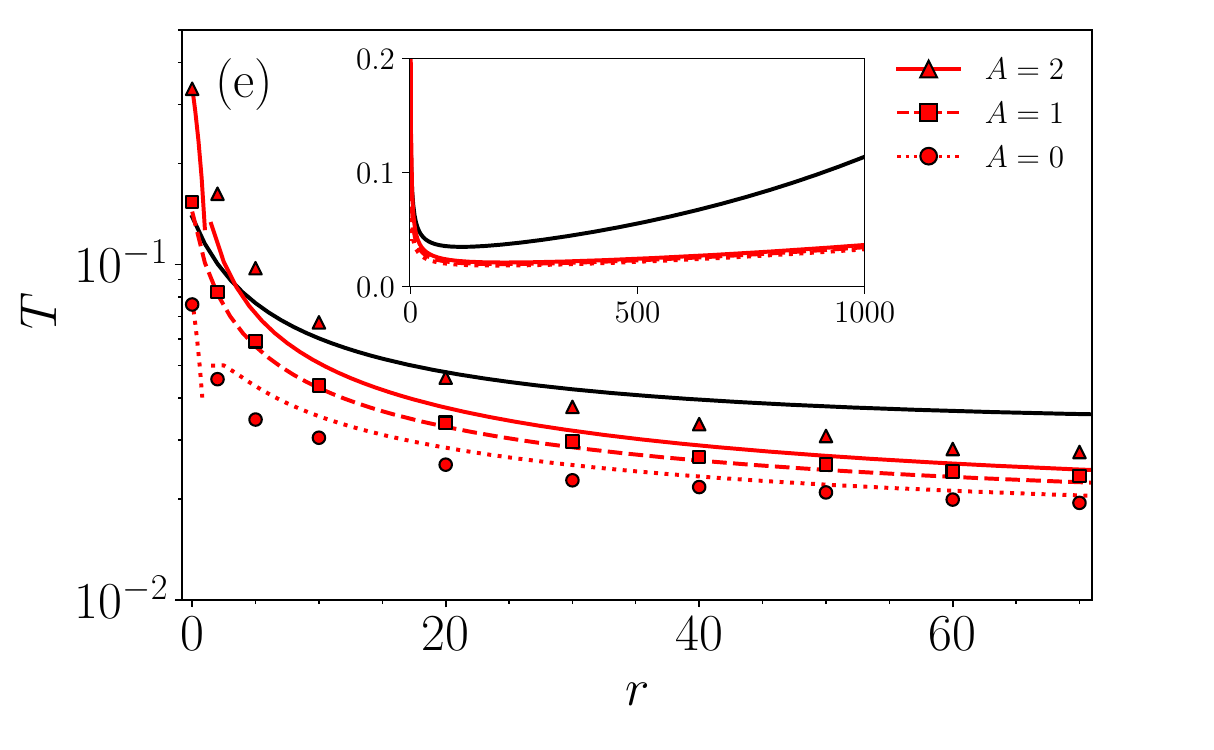}
\includegraphics[width=0.3\textwidth]{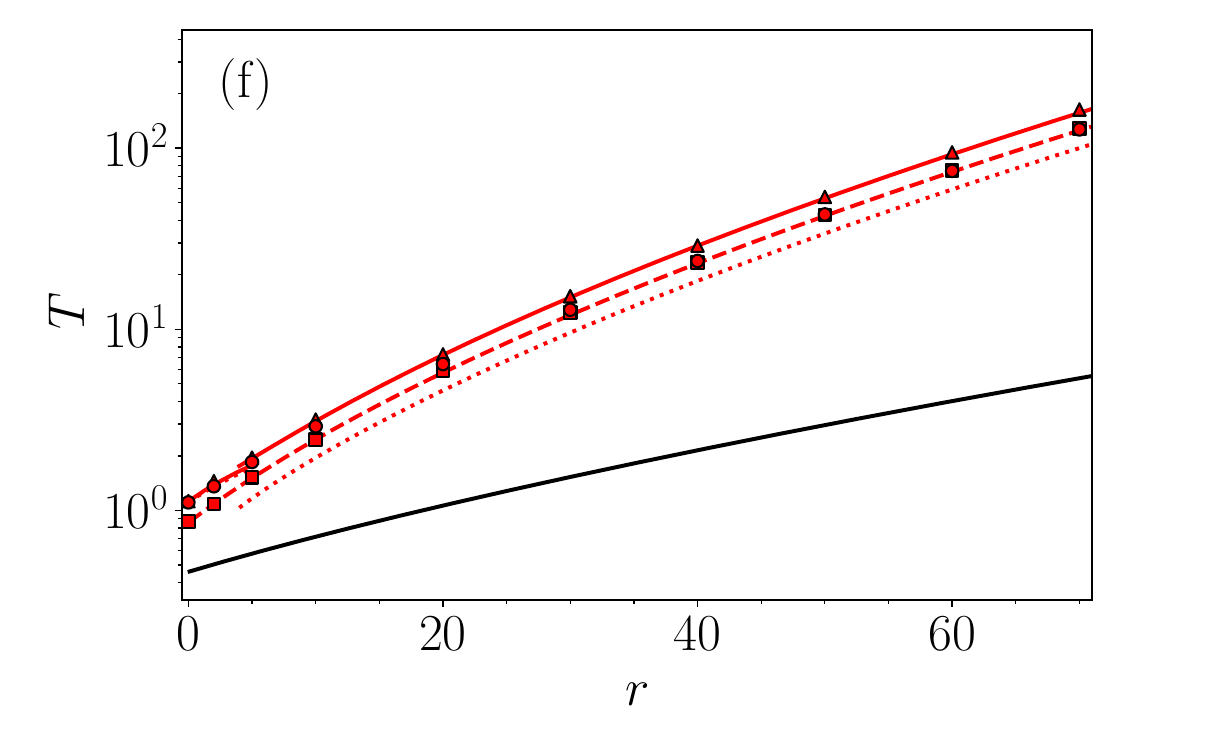}
\caption{ %
(a) and (d): Phase diagrams given by 
${\rm sign}(T_1) \equiv dT/dr|_{r=0}$ vs. $x_0$, for the oscillating diffusivity $D(x)=1+d\cos(2\pi x)$  with $d = -0.9$ (orange) and $d = 0.9$ (red). The curves for $d=0$ (black) are also plotted. The profiles are depicted in the respective insets. 
MFPT vs. $r$ for 
$(d,x_0)=$  $(-0.9,0.15)$ (b) and  $(-0.9,0.71)$ (c), 
$(d,x_0)=$  $(0.9,0.15)$ (e) and  $(0.9,0.71)$ (f). 
The approximations for  small $r$, given by Eq.~(\ref{eq:serie}), 
and  for large $r$, given by Eq.~(\ref{eq:wkb-sol}) are plotted. 
In (e), the inset is a magnification of the main graph, to display the minima. 
The symbols correspond to stochastic simulations of Eq.~(\ref{eq:process1b}). 
}
\label{fig:Vprofile}
\end{figure*}

\subsubsection{Large-$r$ approximation}

 For large values of $ r$, we found (see Appendix section \ref{ap:r-large}): 
 
\begin{eqnarray}
\label{eq:wkb-sol}
    T^l(x_0) = \frac{1}{r\, \mathcal{C}_A(\Delta_r,r,L)} \left[ \left(\frac{D(x_t)}{D(x_r)}\right)^{\frac{A-1}{4}} \mathcal{C}_A(\Delta_t,r,L)  \right.\nonumber\\   -    \left. \left(\frac{D(x_0)}{D(x_r)}\right)^{\frac{A-1}{4}}   \mathcal{C}_A(\Delta_0,r,L)\right],\nonumber\\
\end{eqnarray}
where $y_*\equiv y(x_*)$, $\Delta_*=y_L-y_*$,
%
and the function $  \mathcal{C}_A(y,r,L)$ is defined as 
\begin{eqnarray}
\mathcal{C}_A(y,r,L) &=& (A-1) D'(L)\sinh{\left(\sqrt{r} y\right)}\nonumber \\ &+& 4\sqrt{r D(L)}\cosh{\left(\sqrt{r} y\right)}.
\end{eqnarray}

It is important to emphasize  that, setting $A=1$ in Eq.~(\ref{eq:wkb-sol}), one recovers Eq.~(\ref{eq:A=1}), which is valid for any $r$, within Stratonovich scenario.

\subsubsection{Small-$r$ approximation}

For small $r$,  we consider the series
\begin{equation}
    T^s(x_0) = \sum_{n=0}^\infty r^n T_n(x_0),
    \label{eq:serie}
\end{equation}
where the unperturbed ($r=0$) solution was previously obtained~\cite{menon2023}, namely

\begin{eqnarray} 
T_0(x_0) & = & \int_0^{L} D(x'')^{-\frac{A}{2}}dx''  \int_0^{x_0} D(x')^{-1 + \frac{A}{2}}dx'  \nonumber \\  & - &  \int_0^{x_0} D(x'')^{-1 + \frac{A}{2}}  \int_0^{x''} D(x')^{-\frac{A}{2}}dx' dx'', \nonumber \\
\label{eq:GeneralT_main}
\end{eqnarray}
and the higher-order terms are given by
\begin{equation}
    T^s_n(x_0)= \int_0^L G(x|\xi) D(\xi)^{-\frac{A}{2}}\left[T_{n-1}(\xi) - T_{n-1}(x_r) \right]d\xi,
\end{equation}
where $G(x|\xi )$ is  defined in Appendix \ref{app:asymptotic}, Eq.~(\ref{eq:pert-greenf}).

\subsubsection{Application of asymptotic results}

Let us apply the obtained approximations to study a nonlinear non-monotonic profile, the cosine case  $D(x)=D_0+ d\cos(2\pi x/L)$, 
considered above within the Stratonovich scenario, but now for other values of $A$. 
Results for the optimal values are shown in Fig.~\ref{fig:Vprofile}.
In the expected limits, the asymptotic approximations describe well the MFPT obtained in numerical simulations of Eq.~(\ref{eq:process1b}). 
The full lines for small $r$   represent the (linear) approximation given by Eq.~(\ref{eq:serie}), while for large $r$, the approximation is given by Eq.~(\ref{eq:wkb-sol}) is shown.

From the small-$r$ approximation, we can extract the quantity 
\begin{equation} \label{eq:sign}
    \textrm{sign }[T_1(x_0)] \equiv \left.\frac{dT}{dr}\right|_{r=0},
\end{equation}
where $T_1$ is the second coefficient  of Eq.~(\ref{eq:serie}).
Since this coefficient is the derivative of the MFPT at $r=0$, 
then it indicates  the type of regime. 
If positive, there is no optimal $r$, if negative, an optimal $r$ exists \cite{pal2019landau}, because necessarily  $T$ diverges in the limit $r\to\infty$ (in fact,  in this case the searcher infinitely resets and never reaches the target). This procedure is used and represented in  Fig.~\ref{fig:Vprofile} (panels (a) and (d)).

Regarding the impact of resetting, note that, again, although both profiles have the same mean and visit the same values, the one that has higher  diffusivity  near the target (red lines) is the profile for which resetting is more effective, reducing the time $T$ for sufficiently large $r$, compared to the homogeneous case (Fig.~\ref{fig:Vprofile}(e)). While for the profile that presents low diffusivity near the target, $T$ (even the minimum) becomes larger than in the homogeneous case (Fig.~\ref{fig:Vprofile}(e)).

\section{Other conditions}
\label{sec:other}

In this section we analyze other cases of interest: two absorbing boundaries, reset position different from the initial one ($x_r\neq x_0$), and the limit of a semi-infinite domain.

\subsection{Absorbing boundary conditions}
\label{sec:absorbing}

We also consider the situation with two absorbing boundaries, at $x=0$ and $x=L$. 
This setting corresponds, for instance, to the situation where there is a single target in a domain of size $L$ with periodic boundary conditions (a ring), 
and without loss of generality we can put the target at $x=0$. 
It can also correspond to the search in the region between two targets separated by $L$.

Applying the absorbing boundary conditions to the general solution given by in Eq.~(\ref{eq:general}), in   Appendix \ref{ap:two_abs}, we obtained 
\begin{equation}
    T(y_0) =  \frac{
    {\mathcal W}_A(z_0,z_t) -    
    {\mathcal W}_A(z_0,z_L) + 
    {\mathcal W}_A(z_t,z_L) }
    {r\left({\mathcal W}_A(z_r,z_L) -{\mathcal W}_A(z_r,z_t)    \right)  ,   }
    \label{eq:mfpt_hete_two}
\end{equation}
where,
\begin{equation}
\mathcal{W}_A(u,v) = (uv)^\frac{A}{2}\left(K_{\frac{A}{2}} (u)I_{\frac{A}{2}} (v)- I_{\frac{A}{2}} (u) K_{\frac{A}{2}} (v)\right). 
 \label{eq:wdef}
\end{equation}

\begin{figure}[b!]
\includegraphics[width=0.45\textwidth]{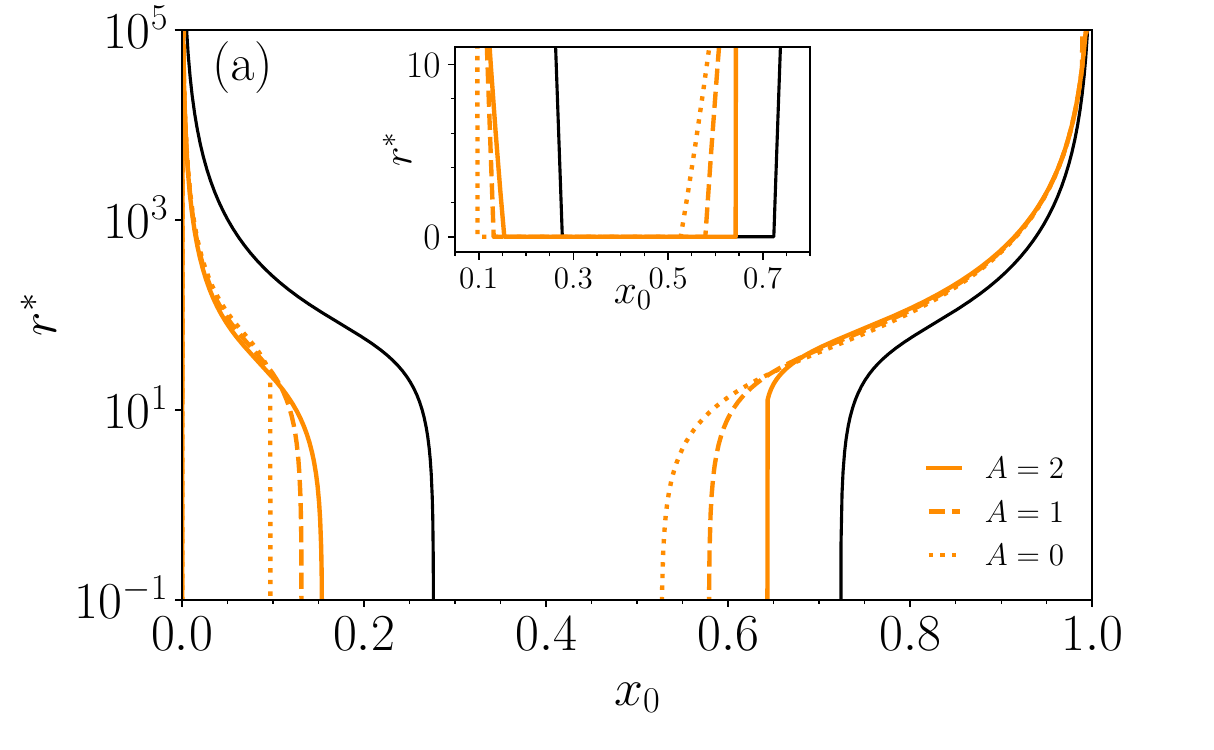}
\includegraphics[width=0.45\textwidth]{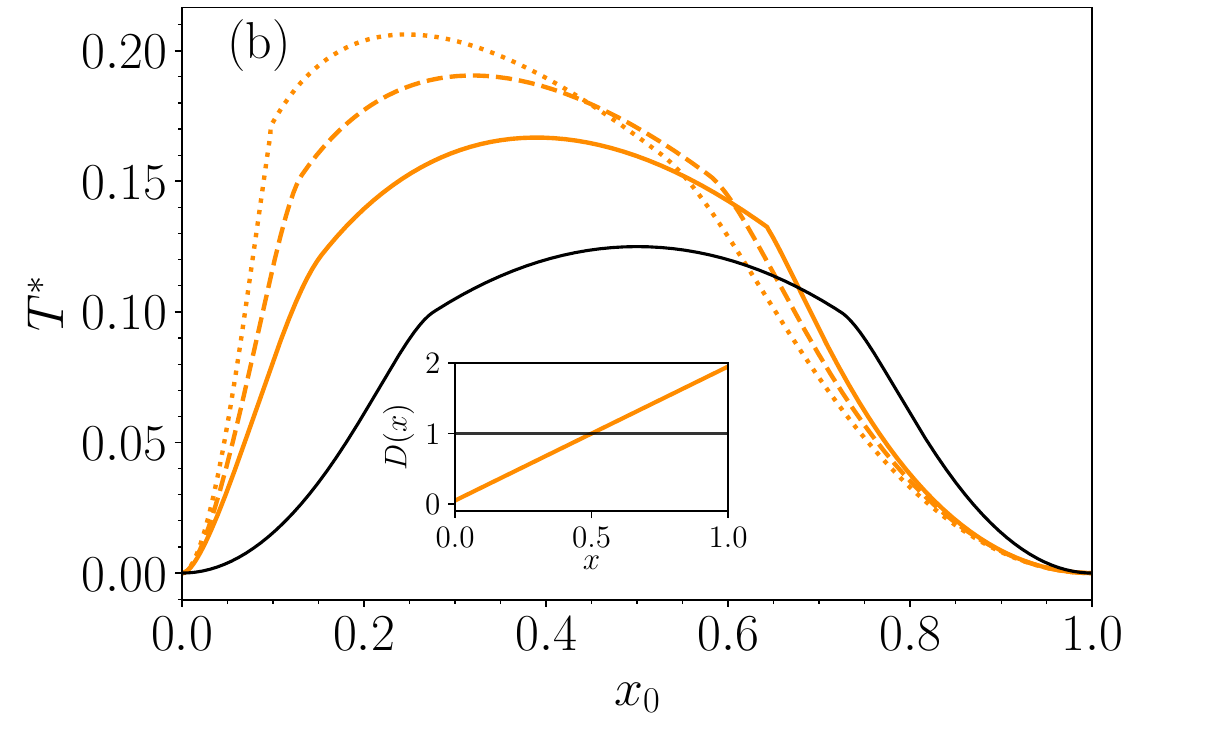}
\caption{{\bf 
Linear diffusivity $D(x)=1+1.9(x-1/2)$,} depicted in the inset of the lower panel, 
with absorbing boundaries at $x=0$ and $x=L=1$
    (a)   Optimal resetting rate $r^*$  vs. $x_0$, for different values of $A$ indicated in the legend. 
    (b) Corresponding optimal MFPT $T^*$ vs. $x_0$.
}
\label{fig:two_absorbing}
\end{figure}

As a control, we note that, in the homogeneous case, Eq.~(\ref{eq:mfpt_hete_two}) becomes

\begin{equation}
\label{eq:mfpt_homo_two}
    T(x_0) = \frac{2 \sinh{\left[ 
\sqrt{r/4D_0}(L-x_0) \right]\sinh{\left[ 
\sqrt{r/4D_0}\,x_0 \right] }}}{r \cosh{\left[ 
\sqrt{r/4D_0}(L-2x_r) \right]}},
\end{equation}
in agreement with Ref. ~\cite{pal2019resetinterval}.

In Fig.~\ref{fig:two_absorbing}, we present $r^*$
and $T^*$ vs. $x_0$ for the linear case with two absorbing boundaries. Due to the symmetry of the borders, we consider only $\alpha>0$ in this case.
Note that, in the vicinity of each border (target position), we observe the same behavior for the linear case with a single absorbing boundary (Fig.~\ref{fig:linearD}), depending on whether the diffusivity is minimal or maximal at the respective targets.

\begin{figure}[b!]
\includegraphics[width=0.32\textwidth]{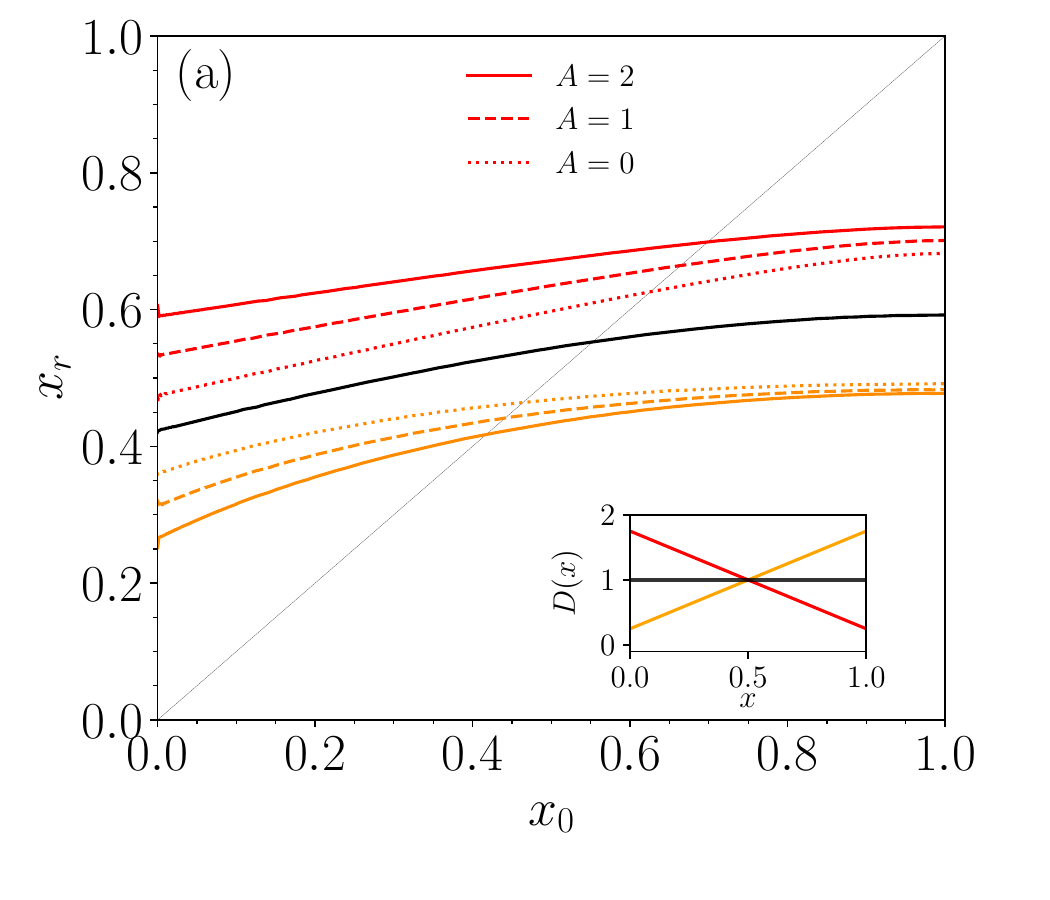}
\includegraphics[width=0.32\textwidth]{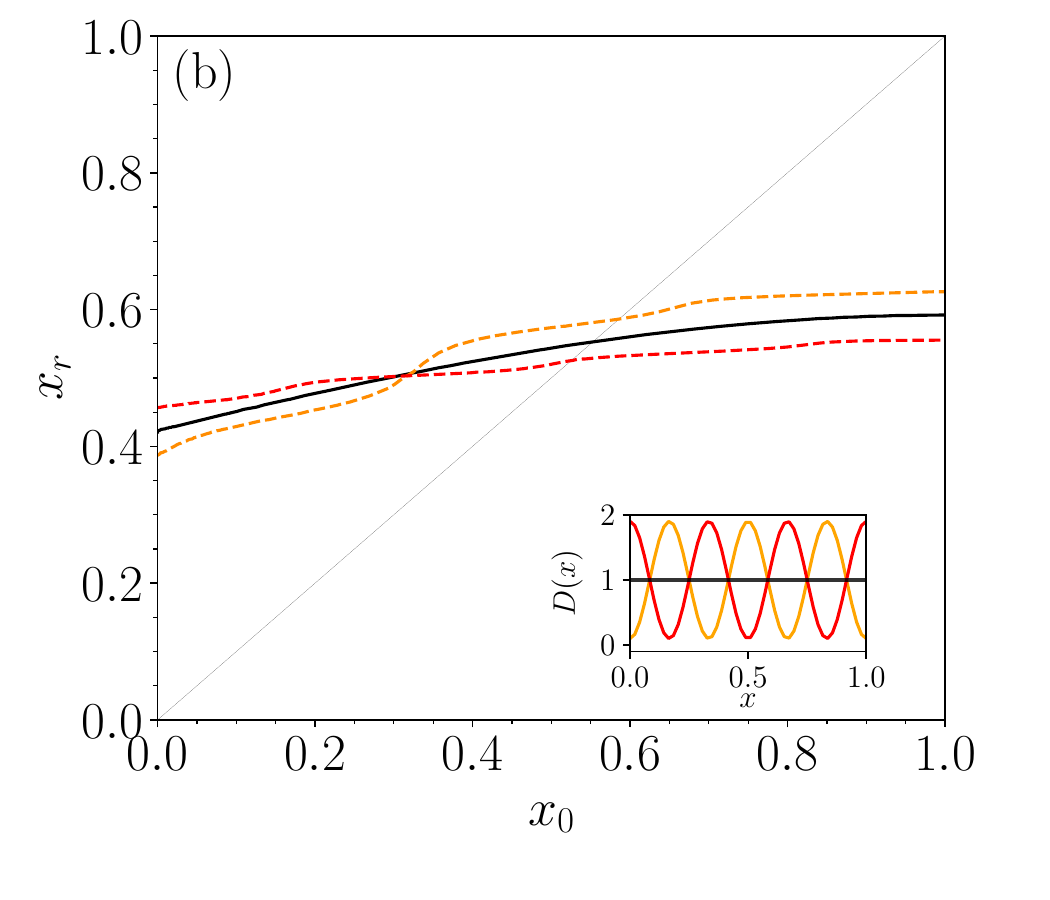}
\caption{Critical lines. 
(a) {\bf 
Linear diffusivity } $D(x)=1+\alpha(x-1/2)$,  
, with $\alpha = 1.5$ (orange) and $\alpha = -1.5$ (red),
different values of $A$ indicated in the legend.
(b) {\bf Oscillating diffusivity} $D(x) = 1 + \pm d\cos\left( 6\pi x \right) $,  for 
    $d = 0.9 $ (orange), $d =- 0.9 $ (red) and $d=0$, with $\alpha = 1.5$ (orange) and $\alpha = -1.5$ (red),
depicted in the respective insets,  for Stratonovich prescription $A=1$.
The region below each curve corresponds to 
$dT/dr|_{r=0}<1$ , otherwise $dT/dr|_{r=0} > 0$, $x_r=x_0$, is plotted for comparison.}

\label{fig:x0xr}
\end{figure}

\subsubsection{Generic reset position  $x_r\neq x_0$}
\label{sec:xrx0}

 In Fig.~\ref{fig:x0xr}, we represent for different situations, the critical curve that separates 
 the regions where  $dT/dr|_{r=0}<0$ (below the curve) for which resetting is favorable  and where  $dT/dr|_{r=0}>0$ (above the curve)  for which resetting is unfavorable. 
 The curves were determined using the quantity defined in Eq.~(\ref{eq:sign}), calculated for  different values of $x_0$ and $x_r$ for the diverse profiles depicted in the respective insets. We note a strong dependency on the reset position $x_r$, while  the dependency on the initial position $x_0$ is weak, expect for very small $x_0$. 
  In fact for small $x_0$, many trajectories can reach the target before the first resetting, otherwise the memory of the initial condition will be rapidly erased since $r^*$ is large. 
 The weak dependence on $x_0$ is also observed in te homogeneous case~\cite{durang2019first}.

\subsection{Limit $ L\to\infty$}
\label{sec:unbounded}
 
  It is worth discussing our results in the limit of a semi-infinite environment ($L\to \infty$), in comparison with previous results in the literature.
In homogeneous environments, for sufficiently large $L$, there is an optimal $r$ for any $x_0$~\cite{christou2015diffusion}. 
For inhomogeneous media, 
in the semi-infinite setting ($L\to \infty$), 
 the MFPT diverges when $r=0$~\cite{Santos2022}, 
 but  it becomes finite for $r> 0$ and the large-$r$ approximation given by Eq.~(\ref{eq:wkb-sol})  becomes 

\begin{equation}
    T^l(x_0) =   
    \frac{ 
   D(0)^{\frac{A-1}{4}} e^{\sqrt{r} y_r}
-D(x_0)^{\frac{A-1}{4}} e^{\sqrt{r}(y_r-y_0)}
}
{r D(x_r)^{\frac{A-1}{4}} }. \\
 \label{eq:Tlarge}
\end{equation}

It is interesting to note, that for  homogeneous diffusivity,  
Eq.~(\ref{eq:Tlarge}) 
recovers the known result~\cite{evans2011diffusion}:
\begin{equation}
  T(x_0)= \frac{1}{r}\left(
  e^{ \sqrt{r/D_0} x_r} -  e^{ \sqrt{r/D_0} (x_r-x_0)} \right),  
\end{equation}  
which is exact for any $r$, not necessarily large.
Eq.~(\ref{eq:Tlarge}) also gives the exact solution for any $r$, 
within the Stratonovich interpretation ($A=1$), as found in Section \ref{sec:stratonovich}. 

Furthermore,  for $A=1$, Eq.~(\ref{eq:Tlarge})  can also be obtained as 
\begin{equation}
T(x_0)= \lim_{s\to 0} \tilde{Q}(x_0, s), 
\end{equation}
with~\cite{Evans2020} 
\begin{equation}
    \tilde{Q} (x_0,s) = \frac{\tilde{Q}_0(x_r, r+s)}{1-r \tilde{Q}_0(x_r, r+s)},
\end{equation}
 obtained using renewal theory, where
\begin{equation}
    \tilde{Q}_0(x_0,s)=\frac{1}{s}\left(1- e^{\sqrt{s} y(x_0)}\right)
    \end{equation}
 is the survival probability calculated without resetting~\cite{Santos2022}.

\section*{Final remarks}
\label{sec:final}

We studied the performance of stochastic resetting in environments with space-dependent diffusivity. 
For arbitrary $D(x)$, we obtained analytical expressions for the MFPT ($T$), which are exact within Stratonovich framework, and approximate for other prescriptions. We also obtained closed-form results for particular forms of $D(x)$ such as linear diffusivity, for arbitrary interpretation. 
In addition, we discussed several extensions such as the situation of two absorbing boundaries, the limit of semi-infinite domain, and the possibility of $x_r\neq x_0$. 

Depending on the characteristics of the heterogeneity, stochastic resetting may be effective for reset positions further away from the target than in the homogeneous case, and 
   the average time $T$ for reaching the target can be shorter. 
   The time $T$ is typically shorter than in the homogeneous case for diffusivities that decay away from the target  (i.e., higher diffusivity close to the target), with stronger effects the more anticipating the process is.
     The opposite occurs for diffusivities that increase away from the target. 
  Furthermore,  within the Stratonovich framework, highly oscillatory profiles degrade efficiency, even more so the higher the wavenumber.

Several extensions are open for future work. One of them is the study of  probability density functions of arrival times. Another possible continuation  is to consider a space-dependent resetting rates, $r(x)$, for example governed by $D(x)$.

\section*{Acknowledgments}
We are grateful to Maike dos Santos for useful discussions. We all acknowledge partial financial support by the 
Coordena\c c\~ao de Aperfei\c coamento de Pessoal de N\'{\i}vel Superior
 - Brazil (CAPES) - Finance Code 001. C.A. also acknowledges partial financial support by 
 Conselho Nacional de Desenvolvimento Cient\'{\i}fico e Tecnol\'ogico (CNPq), 
and Funda\c c\~ao de Amparo \`a Pesquisa do Estado do Rio de Janeiro (FAPERJ).

\bibliography{ref}

\appendix

\section{MFPT for linear profiles
\label{app:mfpt-linear}}

To solve Eq.~(\ref{eq:edofpt}), we begin by introducing the following transformation
\begin{equation}
  \label{eq:tau_trans}
   \tau(y_0)  = \frac{1}{r} \left[1 + r T(y_r)\right] - T(y_0),
\end{equation}
which, substituted into Eq.~(\ref{eq:edofpt}), gives
\begin{equation}
    \label{eq:tau_app}
  \tau''(y_0)+ \frac{1 - A}{y_0+\lambda} \tau'(y_0) - r \,\tau(y_0) = 0,
\end{equation}
where  $\lambda = 2 \sqrt{D_0 - \alpha L/2 }/\alpha$. 
This type of differential equation can be converted into a Bessel equation by means of the ansatz
\begin{equation}
    \tau(y_0) = (y_0+\lambda)^{\frac{A}{2}} f(y_0), 
\end{equation}
which, substituted into   Eq.~(\ref{eq:tau_app}), leads to
\begin{equation}
    f''(y_0) + \frac{f'(y_0)}{y_0+\lambda}  - \left[r + \left(\frac{A}{2(y_0+\lambda)}\right)^2\right] f(y_0) = 0.
\end{equation}
 This equation can be identified as a Bessel equation of order $A/2$ in the variable 
 $\sqrt{r}(y_0+\lambda)$. Its general solution is  expressed as a linear combination of the modified Bessel functions of the 
 first ans second kinds $I_{\frac{A}{2}}$ 
 and  $K_{\frac{A}{2}}$, respectively, as follows
\begin{equation}
    f(y_0) = c_1 K_{\frac{A}{2}}[\sqrt{r}(y_0+\lambda)] + c_2 I_{\frac{A}{2}}[\sqrt{r}(y_0+\lambda)],
\end{equation}
where $ c_1$  and  $c_2$  are constants determined by the boundary conditions.
We can now express the general solution for  $T(y)$  as 
\begin{align}
     &T(y_0) = \frac{1}{r} \left[1 + r T(y_r)\right] \nonumber \\ & -(y_0+\lambda)^{\frac{A}{2}} \left[ c_1 K_{\frac{A}{2}}[\sqrt{r}y_0+\lambda)] + c_2 I_{\frac{A}{2}}[\sqrt{r}(y_0+\lambda)] \right]. \label{eq:general}  
\end{align}
From the reflective boundary condition $\left.\frac{d T(y_0)}{d y_0}\right|_{y_L} = 0$, we obtain
\begin{equation}
    c_1 =  c_2 \frac{I_{\frac{A-2}{2}}[\sqrt{r}(y_L+\lambda)]}{K_{\frac{A-2}{2}}[\sqrt{r} (y_L+\lambda)]}.
\end{equation}
Now, our solution has the form
\begin{eqnarray}
    T(y_0) &=& \frac{1}{r} \left[1 + r T(y_r)\right] \nonumber \\&+& \frac{c_2 \, (y_0+\lambda)^{\frac{A}{2}} U_A\left(\sqrt{r}(y_0+\lambda), \sqrt{r}(y_L+\lambda)\right)}{K_{\frac{A-2}{2}}[\sqrt{r} (y_L+\lambda)]} ,\nonumber\\
    \label{eq:T0prelim}
\end{eqnarray}
where, 
\begin{eqnarray}
    U_A(v,w) = K_{\frac{A}{2}}(v) I_{\frac{A-2}{2}}(w) + I_{\frac{A}{2}}(v) K_{\frac{A-2}{2}}(w), \nonumber
\end{eqnarray}
as defined in Eq.~(\ref{eq:U_def}). 
Finally, from the absorbing condition at the target, $T(0) = 0$,   we get
\begin{equation}
    c_2 = \frac{(1 + r T(y_r)) \lambda^{-\frac{A}{2}} K_{\frac{A-2}{2}}[\sqrt{r}(y_L+\lambda)]}{r U_A\left(\sqrt{r}\lambda,\sqrt{r}(y_L+\lambda)\right)}, 
\end{equation}
which can be substituted in Eq.~(\ref{eq:T0prelim}), and solving the resulting equation self-consistently, leads to Eq.~(\ref{eq:mfpt-r-L}).

\section{Asymptotic approximations }
\label{app:asymptotic}

\subsection{ Large $r$   ($r>>1$)}
\label{ap:r-large}

To obtain the solution of Eq.~(\ref{eq:mfpt-general}), for large values of $r$, we start by using  the transformation introduced by Eq.~(\ref{eq:tau_trans}), and defining the  perturbative parameter   $\varepsilon = 1/\sqrt{r}$. 
So, Eq.~(\ref{eq:mfpt-general}) becomes
\begin{equation}
    \varepsilon^2 D(x_0)  \tau''(x_0) + \varepsilon^2 \left(1 - \frac{A}{2}\right)D'(x_0) \tau'(x_0) -\tau(x_0)=0. 
    \label{eq:diferential-wkb-eps}
\end{equation}
Now we aim to solve this equation, for small values of $\varepsilon$. For this purpose, we consider the WKB expansion 
\begin{equation}
    \tau(x_0) = \exp\left[{\frac{1}{\varepsilon}\sum_{n=0}^\infty \varepsilon^n S_n(x_0) }\right], 
    \label{eq:wkb_series}
\end{equation}
and apply it  into Eq.~(\ref{eq:diferential-wkb-eps}), obtaining  
\begin{align}
     &  D(x)\left[  (S_0')^2  +  \varepsilon (S_0'' +2S_0'S_1' )  
    + \dots 
    \right]\nonumber \\
    &  +\left(1 - \frac{A}{2}\right)D'(x) \left[ \varepsilon S_0' +\varepsilon^2 S_1'+ \varepsilon^3 S_2'
    +\dots\right]=1. \nonumber \\
\end{align}
This yields 
\begin{eqnarray}
    S_0 = \pm\, y(x_0) = \pm \int_0^{x_0} \frac{dx'}{\sqrt{D(x')}},
\end{eqnarray}
and  
\begin{eqnarray}
    S_1 = \frac{A-1}{4} \ln \left[ D(x_0)\right].
\end{eqnarray}
Then, the WKB approximation at first-order reads
\begin{equation}
    T(x_0) = \frac{1}{r} +  T(x_r)   + D(x_0)^{\frac{A-1}{4}}\sum_{m = \pm} c_{m} \exp{\left[m\,\sqrt{r} \,y(x_0) \right]. }
\end{equation}

 Finally, applying the boundary conditions, we arrive at Eq.~(\ref{eq:wkb-sol}). 
 
First note that the  WKB coefficient $S_0$ naturally leads to the change of variables  defined in Eq.~(\ref{eq:y}) and used  to transform Eq.~(\ref{eq:mfpt-general})  into an homogeneous diffusion equation with the addition of a force term, i.e., Eq.~(\ref{eq:edofpt}).

Furthermore, 
 when $A=1$ (Stratonovich), the coefficient $S_1$ vanishes, and actually all the upper-order ones, in which case   
 the exact solution given by Eq.~(\ref{eq:ts}) is recovered,   
 meaning that  the  WKB result is exact for $A=1$.

\subsection{Small $r$   ($r <<1$)  }
We seek a solution for Eq.~(\ref{eq:mfpt-general})   in the form of a perturbative series of the form
\begin{equation}
    \label{eq:serie-app}
    T(x_0) = \sum_{n=0}^\infty r^n T_n(x_0).
\end{equation}
The zeroth-order term $T_0(x_0) $ satisfies the following differential equation
\begin{equation}
    D(x_0)^\frac{A}{2} \left[D(x_0)^{1-A/2} T'_0(x_0) \right]' + 1 = 0,
\end{equation}
whose  explicit solution is given by \cite{menon2023}
\begin{eqnarray} 
T_0(x_0) & = & \int_0^{L} D(x'')^{-\frac{A}{2}}dx''  \int_0^{x_0} D(x')^{-1 + \frac{A}{2}}dx'  \nonumber \\  & - &  \int_0^{x_0} D(x'')^{-1 + \frac{A}{2}}  \int_0^{x''} D(x')^{-\frac{A}{2}}dx' dx'' . \;\;\;\;
\label{eq:GeneralT_apend}
\end{eqnarray}

To obtain the high-order terms $T_n(x_0)$, we substitute 
Eq.~(\ref{eq:serie-app}) into Eq.~(\ref{eq:mfpt-general}), 
obtaining the recursion relation  
\begin{equation}
 D(x_0)^{\frac{A}{2}}   \left[D(x_0)^{1-A/2}T'_n(x_0) \right]'  =  T_{n-1}(x_0)- T_{n-1}(x_r),
\end{equation}
which allows to obtain  the $n$th  coefficient
\begin{equation}
    T_n(x_0)= \int_0^L G(x|\xi) D(\xi)^{-\frac{A}{2}}\left[T_{n-1}(\xi) - T_{n-1}(x_r) \right]d\xi,
\end{equation}
where the Green function $G(x|\xi)$  obeys
\begin{equation}
\left[D(x_0)^{1-A/2} G'(x_0|\xi) \right]'  =  \delta(x_0 - \xi),
\end{equation}
and the explicit form of the solution is
\begin{equation}
\label{eq:pert-greenf}
    G(x_0|\xi)= 
    -\left\{
    \begin{array}{ll}
        \int_0^{x_0} D(x')^{A/2 -1} \, dx' & \text{if }\; 0 \leq x_0 \leq \xi, \\[2mm]
        \int_0^\xi D(x')^{A/2 -1} \, dx' & \text{if }\; \xi \leq x_0 \leq L,
    \end{array}
    \right.
\end{equation}
that verifies the boundary conditions of the problem.

\begin{figure}[b!]
    \centering
\includegraphics[width=0.45\textwidth]{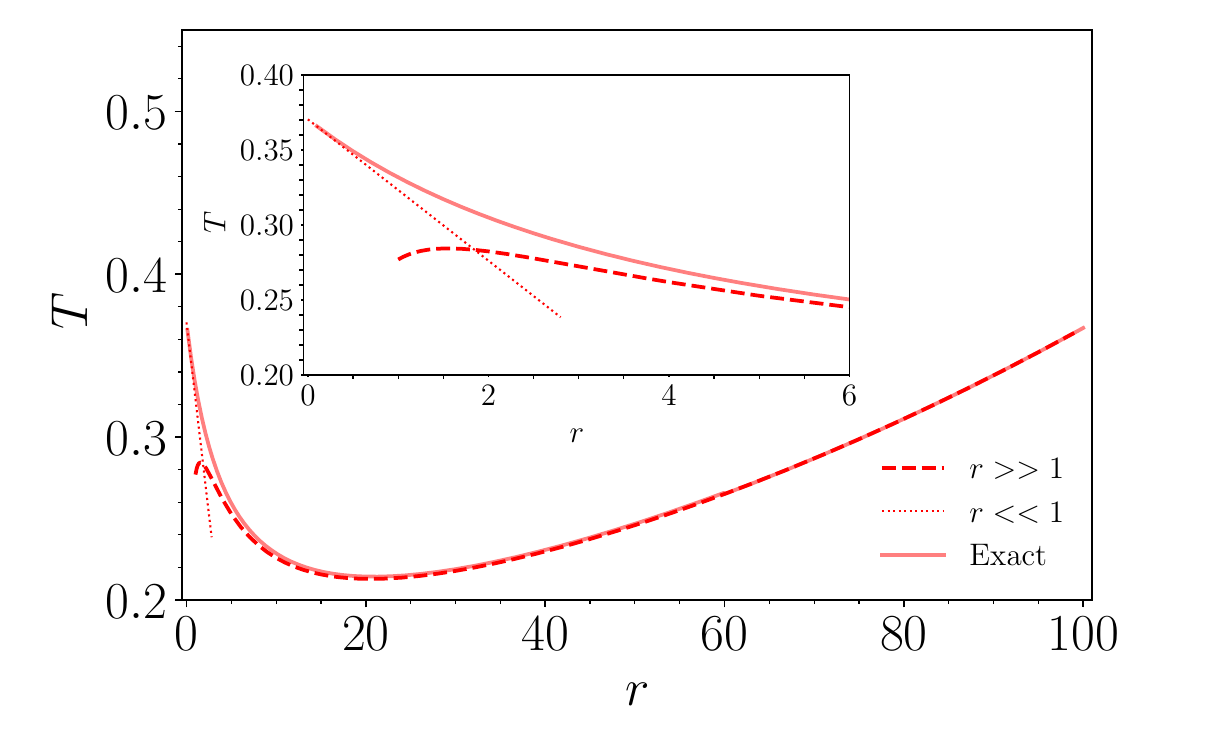}
\caption{ {\bf Linear profile }
$D(x) = 1 -0.9(x - 1/2)$,  within It\^o interpretation ($A=2$). 
MFPT $T$ as a function of the resetting rate $r$.
The asymptotic approximations, 
 for small $r$  given by Eq.~(\ref{eq:serie}) 
 and  for large $r$ given by Eq.~(\ref{eq:wkb-sol}), 
are shown, in good agreement with the exact solution given by Eq.~(\ref{eq:mfpt-r-L}).
The inset shows a magnification of the cross-over. 
} \label{fig:comparison}
\end{figure} 
The good performance of these approximations is illustrated in Fig.~\ref{fig:comparison}, for a linear profile, under 
 It\^o prescription.

\section{Two absorbing boundaries}
\label{ap:two_abs}
The setting with two absorbing boundaries at $x=0$ and $x=L$ can be solved using the general solution given by in Eq.~(\ref{eq:general}). Applying  the boundary condition $T(y_t) = 0$, we get 
\begin{equation}
    c_1 = \frac{\left(\frac{\sqrt{r}}{z_t}\right)^{A/2} \left[ 1+ r T(y_r)  - c_2\left(\frac{\sqrt{r}}{z_t}\right)^{A/2} I_{A/2}(z_t) \right]}{r K_{A/2} (z_t)} , 
\end{equation}
recalling that  we defined $z_*\equiv \sqrt{r}(y_*+\lambda)$. 

From the condition $T(y_L)$=0, we obtain $c_2$, namely
\begin{equation}
    c_2 = \frac{(1+r T(y_r))\left[K_{\frac{A}{2}}(z_L) \left(\frac{\sqrt{r}}{z_t}\right)^{A/2}- K_{\frac{A}{2}}(z_t ) \left(\frac{\sqrt{r}}{z_L}\right)^{A/2} \right]}{r W_A\left(z_L,z_t\right) }, 
\end{equation}
where $  
   W_A(u,v) = K_{\frac{A}{2}}(u) I_{\frac{A}{2}}(v) -  K_{\frac{A}{2}}(v) I_{\frac{A}{2}}(u)$.
 
Using the  coefficients $c_1$ and $c_2$, and solving self-consistently Eq.~(\ref{eq:general}), we arrive at 
Eq.~(\ref{eq:mfpt_hete_two}).

\end{document}